\documentclass[epj]{svjour}
\newcommand{\beq}{\begin{equation}}
\newcommand{\eeq}{\end{equation}}
\newcommand{\bea}{\begin{eqnarray}}
\newcommand{\eea}{\end{eqnarray}}
\newcommand{\uvec}[1]{{\bf \hat{#1}}}
\newcommand{\eq}[1]{{Eq. (\ref{#1})}}

\usepackage{graphics}
\usepackage{dcolumn}
\usepackage{graphicx}
\usepackage{graphics}
\begin{document}
\title{Spin breather and rogue formations in a Heisenberg ferromagnetic system}
\author{Rahul O. R.\inst{1} \and S. Murugesh\inst{1}\fnmsep\thanks{\email{murugesh@iist.ac.in}}
}                     
\offprints{}          
\institute{Department of Physics, Indian Institute of Space Science
  and Technology, Thiruvananthapuram-695~547, India.}
\date{Received: date / Revised version: date}
%
\abstract{We construct explicit spin configurations for the breather solution of a one-dimensional Heisenberg ferromagnetic 
spin system.  This corresponds to the breather soliton solution of the gauge equivalent nonlinear Schr\"{o}dinger equation. There are three
broad cases, wherein the solution shows distinct behavior.
Of particular interest to our study is the rogue behavior in spin field terms. 
} 
\maketitle
\section{Introduction}
\label{intro}
The Heisenberg model of spin interaction has been a subject of keen interest for decades. Although 
fundamentally quantum in nature, for low level excitations its classical 
counterpart has nevertheless yielded important physically realizable results. The bare Heisenberg
model in one-dimension (1-d) is perhaps the simplest yet non-trivial nonlinear 
model among the larger class of O(3) sigma models\cite{rr:1987,zakr:1989}. The 
1-d model is completely integrable with soliton solutions, and is 
gauge equivalent to the nonlinear Schr\"{o}dinger equation (NLSE)\cite{ml:1977,takt:1977,zs:1972}. The classical model in 2-d, though not known to be 
integrable, has a very interesting class of solutions called
skyrmions that fall into distinct topological sectors identified by the Hopf invariant\cite{bp:1975}.

The NLSE is another fundamental model naturally and frequently arising in a 
variety of physical systems such as fluid dynamics, dynamics of
polymeric
fluids, fiber optics and vortex dynamics in superfluids, to name a few 
\cite{hb:1982,hasi:1972,bird:1987,hase:1989,schw:1988}. Besides its physical importance, it also is a very important
model in soliton theory, owing to its rich mathematical structure. It is 
completely integrable with soliton solutions, and presents itself amenable to nearly
every method available in the study of nonlinear systems, making it a perfect
pedagogical model. Its complete integrability was first established in
a classic paper by Zakharov and Shabat in 1972, which also brought
about a deeper clarity, from a geometric point of view, on the method of 
inverse scattering transform being developed around that period\cite{zs:1972,sulem:1999}. For this reason it remains one of the most studied, and among
the most understood of nonlinear systems. Besides NLSE, further extensions of 
the system, such as the N-coupled NLSE, multidimensional NLSE, higher order 
NLSE, to name a few, have often proved fruitful in studying several physical 
systems of importance. Inspite of its rich history and continued interest, investigations on NLSE have often thrown out novel results and physical behavior never anticipated or intuited earlier. One such is the rogue wave solutions,
marked by a sudden, momentary yet colossal enhancement in 
the field variable\cite{ma:1979,pere:1983,akhm:1986}. 
Interest in this special solution have grown manifold since citing of
similar
phenomenon in the deep ocean\cite{kharif:2008}. Besides, the same have
been
predicted to occur in Bose-Einstein condensates, whose dynamics is
modeled by the Gross-Pitaevskii equation, which closely resembles the
NLSE in higher dimension\cite{akhm:2009}.

In this paper, we exploit
this gauge equivalence between the Heisenberg ferromagnet 
and NLSE to study and illustrate the equivalent rogue behavior 
in an 1-d spin field. Furthermore, a natural
connection exists between the spin field in 1-d and 
a space curve in 3-d, its dynamic evolution spanning a surface. In the case of systems integrable and endowed with a
Lax pair the {\it soliton surfaces} thus obtained are of much
interest geometrically in soliton theory\cite{sym:1985}.  Often the space curves thus obtained are by themselves  
physically realizable, adding to their
significance\cite{sm:2005}. Here, however, we
shall restrict ourselves to obtaining the breather spin field for the
Heisenberg spin chain, especially demonstrating the rogue behavior. In
the next section, we introduce the model briefly and, after an outline
of the method of obtaining the solution, give the general
expression for the breather spin field. We also discuss in detail
two cases, determined by the relative values of the relevant
parameters, wherein the behavior of the breather modes are 
pronouncedly distinct. The rogue phenomenon is illustrated as a
special limiting 
case of these two behaviors. 

\section{The rogue spin field}
\label{sec:2}
The classical Heisenberg ferromagnet (HF) is governed by the hamiltonian
\beq
\mathbf{H} = J\int{\nabla\uvec{S}}^2
\eeq
where $J$ is a constant arising out of the exchange integral,
positive for a ferromagnet, and $\uvec{S}(x,t)=\{S_1,S_2,S_3\}$ is the unit spin field ($|\uvec{S}|=1$)
of interest. In this paper, we shall be concerned only with
the 1-d spin chain. Using the spin commutation relations one gets the simplest form of the Landau-Lifshitz equation (LLE) governing its dynamics (after rescaling $t$)
\beq\label{ll}
\uvec{S}_t  = -\uvec{S}\times\uvec{S}_{xx},
\eeq
where the subscripts stand for partial derivatives, as usual. As 
mentioned earlier, the 1-D LLE, \eq{ll}, is integrable through inverse
scattering method with soliton solutions, and can be expressed
through the auxiliary linear equations\cite{fad:1987}
\bea\label{lap}
\Phi_x = U_{HF}\Phi = \frac{-i\lambda}{2}{\bf S}\Phi\nonumber\\
\Phi_t = V_{HF}\Phi = \Big{(}\frac{\lambda}{2}{\bf S}_x{\bf S} +\frac{i\lambda^2}{2}{\bf S}\Big{)}\Phi.
\eea
Here
\beq\label{spin}
{\bf S} = S_i\sigma_i,
\eeq
(summation assumed) with $\sigma_i,\,\,i=1,2,3$ being the Pauli matrices, and $S_i$ the components of the unit spin field $\uvec{S}$, and $\lambda$ is the scattering parameter.
Thus, $U_{HF}$ and $V_{HF}$ form the Lax pair for the system.  The compatibility of the two equations,
\eq{lap}, gives the matrix form of the LLE:
\beq
{\bf S}_t = \frac{-i}{2}[{\bf S}, {\bf S}_{xx}],
\eeq
which, when expressed in scalar form, gives back \eq{ll}.

The most prominent solution of the Heisenberg model in 1-d
is the localized traveling wave solution of the
`secant-hyperbolic' variety, associated with the {\it isolated} soliton solution of the NLSE.  
 The NLSE
\beq
i\psi_t+\psi_{xx}+2|\psi|^2\psi=0
\eeq
supports the trivial solution $\psi=0$. Taking this as a seed solution, any of the standard methods of obtaining soliton
solutions would yield a one-soliton solution of the form 
\beq\label{1sol}
\psi = \lambda_{0I}{\rm sech}\omega_{0I}e^{i\omega_{0R}},
\eeq 
where 
$\omega_0=\omega_{0R}+i\omega_{0I}= \lambda_0x-\lambda_0^2t-(\varphi_1+i\varphi_2)$, and 
$\lambda_0=\lambda_{0R}+i\lambda_{0I}$ is the
complex eigenvalue on the plane of the scattering 
parameter that determines the amplitude and velocity
of the soliton. $\varphi_{1,2}$ are arbitrary constants indicating
the initial phase and position of the soliton. Of particular interest to us for the
following discussion, as we shall see below, is when $\lambda_0$ 
is purely imaginary, in which case the soliton is
stationary, localized spatially and temporally periodic. 

The NLSE also permits a temporal wave solution of the
form 
\beq\label{seed}
\psi_0(x,t) = Ae^{2iA^2t}, 
\eeq
where $A$ is some {\it positive real} constant. While
the isolated soliton corresponds to a vanishing boundary
condition at infinity, this seed form, \eq{seed}, leads to a solution that decays 
to a uniform solution. 
Starting with this seed solution (or equivalently, a spatially 
static wave solution of the form $\psi_0 = ke^{\pm i\frac{k}{\sqrt{2}}x}$, $k$ being any real constant, or a plane traveling wave with a suitable dispersion relation, obtained through an appropriate gauge 
transformation to $\psi(x,t)$), one may still go ahead to 
find a more non-trivial solution through any of the
standard inverse scattering transform methods. Such a solution, {\it the breather},  has indeed 
been obtained by a host of	 authors\cite{ma:1979,pere:1983,akhm:1986}.  The general form of the solution is given by
\beq\label{rogue}
\psi_r = e^{2 i A^2 t} \Bigg[ A\, +\,  2\, \lambda_{0I}\, \frac{\big(\zeta - i\, \eta \big)}{Z} \, \Bigg]
\end{equation}
where $\lambda_0$ is the complex eigenvalue, as remarked earlier, associated with the one-soliton solution, and 
$ A $ is the amplitude in \eq{seed}. \\
$Z = -2\, A\, \Im(\mu)\, \cos(2\, \Re(\Omega_0)) + \alpha\, \cosh(2\, \Im(\Omega_0) + 2K)  $\vspace{10pt}\\
$\zeta = \alpha\, \cos(2\, \Re(\Omega_0)) - 2\, A\, \Im(\mu)\, \cosh(2\, \Im(\Omega_0)+ 2K)
$\\
$\eta = -\beta\, \sin(2\, \Re(\Omega_0)) + 2\, A\, \Re(\mu)\, \sinh(2\, \Im(\Omega_0)+ 2K)
$\\
$\Omega_0 = 2\, t\, p_0\, \lambda_0\, + p_0\, x\, - \phi_0$\\
$p_0^2 = A^2 + {\lambda_0}^2 $. \\
$\mu = ( \lambda_0 - p_0 )$ \\
$\alpha = 2\, \Big( \lambda_{0I}^2 + \Re(p_0)^2 - \lambda_{0R}\,  \Re(p_0) - \lambda_{0I}\, \Im(p_0) \Big)$\\
$\beta = 2\, \Big( \Im(p_0)^2 + \lambda_{0R}^2 - \lambda_{0R}\,  \Re(p_0) - \lambda_{0I}\, \Im(p_0) \Big )$\\
For special values of $A$ and $\lambda_0$, it displays a
{\it rogue} behavior, marked by sudden, short, yet colossal
rise in the field magnitude. 

Here we present its spin analogue - {\it the rogue spin wave}. 
To do so, we briefly outline the procedure to be followed using the gauge equivalence of the 1-d Heisenberg model and the
NLSE \cite{takt:1977,fad:1987,schief:2002}.
It may be noted that the spin field ${\bf S}$, as expressed
in \eq{spin}, and the Lax pair in \eq{lap} are  elements of the $su(2)$ lie algebra. 
Consequently, the auxiliary function $\Phi$ lives in the
$SU(2)$ group. The auxiliary equations associated with the
NLSE similarly indicate that the Lax pair for NLSE are
elements of $su(2)$, with auxiliary functions in the 
$SU(2)$ group. If we denote these auxiliary functions by $\Phi_{NLS}$
then the spin field ${\bf S}$, \eq{spin}, and the equivalent LLE, \eq{ll}, can be obtained through a unitary transformation
\beq
{\bf S} = \lim_{\lambda\to 0}\Phi_{NLS}^{\dagger}\sigma_3\Phi_{NLS}^{~}.
 \eeq
 Finally, one writes 
 the corresponding vector field $\uvec{S}$ making  use of the isomorphism between $su(2)$ lie algebra and the euclidean $\mathbf{R}^3$. The 
 energy densities in the two descriptions are related as
\beq
\mathbf{E}=\frac{1}{2}\uvec{S}_x^2=|\psi|^2,
\eeq
and the momentum density of the spin field is given
by\cite{tjon:1977}
\beq
\mathbf{p}_x= \frac{\partial\phi}{\partial x}(1-S_3),
\eeq
where $\phi=\tan^{-1}(S_2/S_1)$. 

Going by this procedure, the spin field corresponding to the seed solution \eq{seed} 
is derived to be
\bea\label{s0}
\uvec{S}_0= \tanh(2K)\sin (2px-2\phi_0)\uvec{i}
+\cos (2px-2\phi_0)\uvec{j}\nonumber\\
+ {\rm sech}(2K)\sin (2px-2\phi_0)\uvec{k},
\eea
where $p=|A|$, and 
$K$ and $\phi_0$ are arbitrary constants,
arising out of the translational invariance and global
rotation symmetry of LLE, respectively.
It may be noted that this seed spin field is independent
of $t$, but depends on $x$ (although $\psi_0$ is not), and provides a simple 
yet non-trivial static solution for LLE, \eq{ll}.
Besides, it is periodic in $x$. These two observations
aren't surprising. For, as indicated earlier, it merely
amounts to a suitable gauge transformation to the
NLSE seed solution $\psi_0(x,t)$, \eq{seed}.   

Continuing with this procedure to find the one-soliton
solution, after some straight forward but tedious calculation, the 
rogue spin wave is found to be of the form:
\bea\label{s1}
\uvec{S}_1=  \frac{p}{A|\lambda_0|^2} \big(\lambda_{0R}^2 - \lambda_{0I}^2 \big)\uvec{S}_0
+\Big{[}- \frac{2\, \lambda_{0I}^2}{A\, Z^2 |\lambda_0|^2}\Pi\xi\nonumber\\
+ \frac{2}{A\, Z|\lambda_0|^2	}\, \lambda_{0I}\, \lambda_{0R}\, \Big(
A\, \tanh(2K)\, \cos(2px-2\phi_0)\, \eta \nonumber \\
- p\, {\rm sech}(2K)\, \zeta \Big)\Big{]}\uvec{x} \nonumber\\
+ \Big{[}\frac{2\, \lambda_{0I}^2}{A\, Z^2|\lambda_0|^2 }\, \Delta\, \xi
- \frac{2}{ Z|\lambda_0|^2	}\, \lambda_{0I}\, \lambda_{0R}\, \sin(2px-2\phi_0)\, \eta\Big{]}\uvec{y} \nonumber \\
+ \Big{[}\frac{2\, \lambda_{0I}^2}{A\, Z^2|\lambda_0|^2 } \ \Gamma\, \xi\nonumber\\
		- \frac{2}{A\, Z|\lambda_0|^2	}\, \lambda_{0I}\, \lambda_{0R}\, \Big(
	A\, {\rm sech}(2K)\, \cos(2px-2\phi_0)\, \eta 
	\nonumber \\
	+ p\, \tanh(2K)\, \zeta \Big)\Big{]}\uvec{z},~~~~~~
	\eea
where, besides the parameters defined below \eq{rogue},\\
$\Gamma = A\, {\rm sech}(2K)\, \cos(2px-2\phi_0)\, \zeta
-p\, \tanh(2K)\, \eta\\
~~~~~+p\, {\rm sech}(2K)\, \sin(2px-2\phi_0)\, \xi$\\
$\Pi = -p\, \tanh(2K)\, \sin(2px-2\phi_0)\, \xi\\
 ~~~~~-A\, \tanh(2K)\, \cos(2px-2\phi_0)\, \zeta
-p\, {\rm sech}(2K)\, \eta$\\
$\Delta = p\, \cos(2px-2\phi_0)\, \xi
- A\, \sin(2px-2\phi_0)\, \zeta$\\
$\xi = -2\, A\, \Re(\mu)\, \sin(2\, \Re(\Omega_0))- \beta\, \sinh(2\, \Im(\Omega_0)+ 2K)$\\
The form and behavior of spin field is dictated by two 
parameters --- the amplitude $A$, and $\lambda_0$ the complex eigenvalue on
the plane of the scattering parameter associated with
the one soliton. The relative values
of the two parameters lead to broadly three characteristic
behavior in the spin field. However, as is clear from the form of the solution \eq{s1}, it is cumbersome to analyze the 
behavior in its entire generality. For our further analysis,
therefore, 
we shall consider only the special case $A>0$, and that the eigenvalue is purely
imaginary, $\lambda_0=ia$, where $a$ is positive
(It may be recalled that for the NLSE the eigenvalues 
can assume values only on the upper half of the complex plane of the scattering parameter\cite{spn:1984}).  The spin field in \eq{s1} then takes a simpler form
\bea
\uvec{S}_1= - \frac{2\xi}{A\, Z^2 }\Big{[}\Pi \uvec{x} 
- \Delta\, \uvec{y} -\Gamma\, \uvec{z}\Big{]} - \uvec{S}_0.
\eea

There are three distinct cases i) $A<a$, ii) $A>a$  iii) $A=a$. However, the last case is trivial, reducing to the
background field due to the seed. The more interesting
situation is $A \sim a$ which leads to a rogue
behavior. We discuss these three situations  in 
detail below:

\subsection{Case i: $A<a$}
Firstly, it is verified that as $A\to 0$, the form of the rogue
soliton, \eq{rogue}, approaches the isolated one-soliton, \eq{1sol},
as expected. An illustrative plot of the energy and momentum densities for $A(\ne 0)<a$, are plotted in Fig. 1. 
The spin field is localized in space, but shows an  oscillatory behavior in time, with a period 
\beq\label{period1}
T=\frac{\pi}{2a\sqrt{a^2-A^2}}.
\eeq
As remarked earlier, this behavior is that of the isolated one-soliton for purely imaginary
$\lambda_0$. The form of the spin field itself is shown in
Fig. \ref{field1}.a. 
\begin{figure}
	\resizebox{0.5\textwidth}{!}{\includegraphics{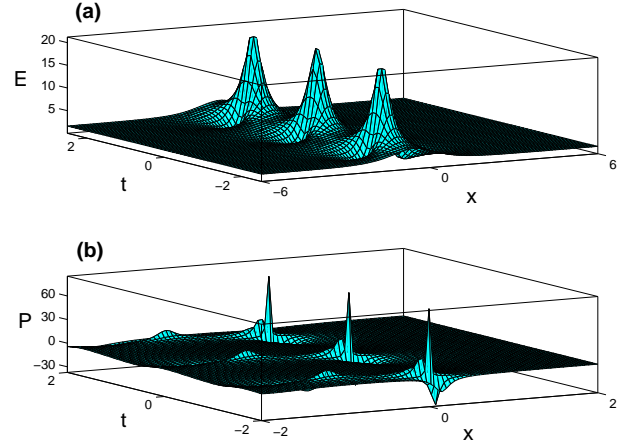}}
	\caption{The energy (a) and momentum (b) densities for the
          temporally periodic, and spatially localized, spin field in
          the $x-t$ plane for $A<a$ --- Kuznetsov-Ma breather. Qualitatively, this is similar to
          the one-soliton spin field for the 1-D HF model, except that
          it is non-traveling, due to the eigenvalue $\lambda_0$ being
          purely imaginary.} 
\end{figure}
\begin{figure*}[t]
	\resizebox{1\textwidth}{!}{\includegraphics[clip,trim=0cm 3cm 0cm 3cm,scale=1.0]{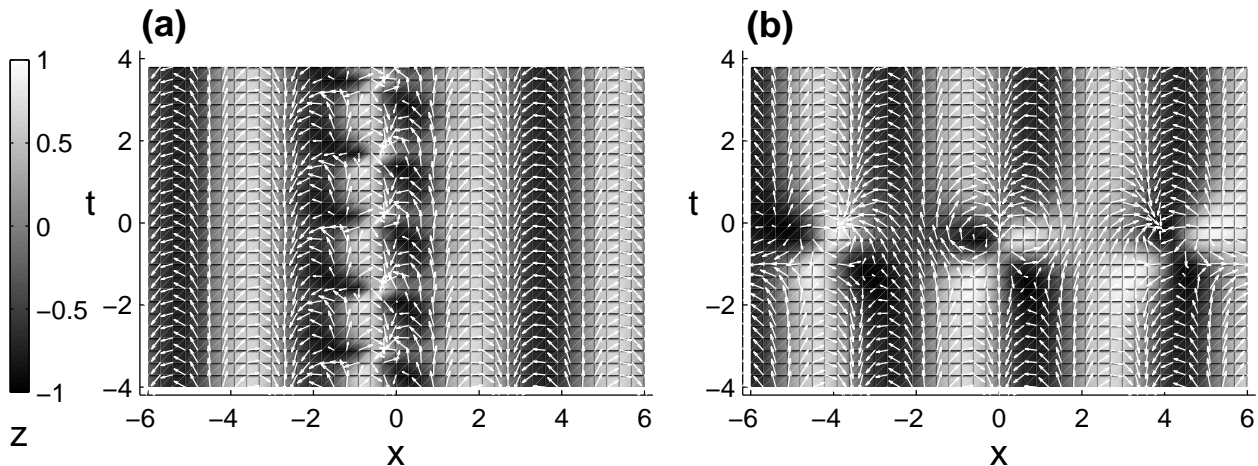}}
	\caption{The spin fields (2-d projections) for the two cases i. $A<a$ --- Kuznetsov-Ma breather (a) and ii. $A>a$ --- Akhmediev breather (b). Grey scale indicates the strength of the $z$-component, $S_z$.} 
	\label{field1}
\end{figure*}

\subsection{Case ii: $A>a$}
\begin{figure}
	\resizebox{0.5\textwidth}{!}{\includegraphics{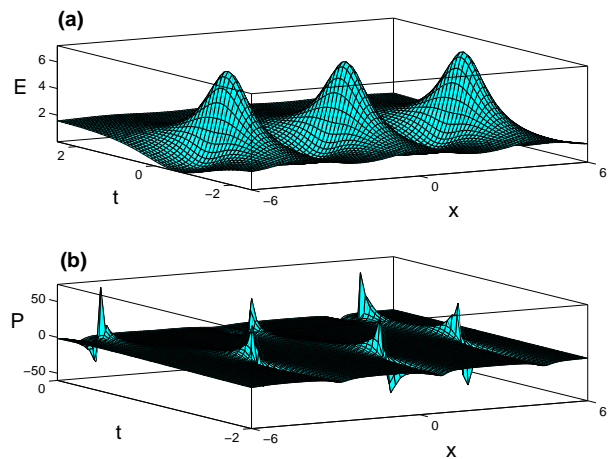}}
	\caption{The energy density $\mathbf{E}$ (a) and the momentum
          $\mathbf{P}$ (b) fields in the $x-t$ plane for $A>a$ --- Akhmediev breather. The
          periodicity in time, seen in case i, is replaced by a
          spatial periodicity. Temporally the excitations are localized.} 
	\label{c2energy}
\end{figure}

At $A=a$,  the spin field is
reduced to the background seed field that we started
with. 
However, as the magnitude of $A$ makes the cross over from
$A<a$ to $A>a$, we see an interesting change in
the form of the associated spin field. The temporal
periodicity witnessed earlier is replaced by a spatial
periodicity, while more interestingly, the field is now
temporally local. Spatially, the field is composed of
two periodic functions with periods
\beq\label{period2}
L_1 = \frac{\pi}{A},\,\, {\rm and}\,\, L_2 = \frac{\pi}{\sqrt{A^2-a^2}},
\eeq
which are in general non-commensurate.
 Neither of the two field configurations presented here possess a  traveling nature owing to the
fact that we have chosen $\lambda_0$ to be purely
imaginary for the sake of ease in analysis. A more
general complex $\lambda_0$, while retaining the
overall general property, renders the filed a traveling
nature with uniform speed determined by $\lambda_0$, 
as is expected of a soliton.  The energy and momentum densities
are shown in Fig. \ref{c2energy}, while the field configuration is shown in Fig. \ref{field1}.b. 

\subsection{Case iii: $|A-a|=\epsilon\sim 0$ --- the rogue field}
We do not discuss the case $A=a$ in detail, as, in that  case the spin field is 
reduced to the seed field, \eq{s0}, i.e, the background field. 
The field is time independent, periodic in space and has a uniform energy density. 

\begin{figure}[h]
    	\resizebox{0.6\textwidth}{!}{\includegraphics[clip,trim=0cm 0cm 0cm 0cm,scale=1.0]{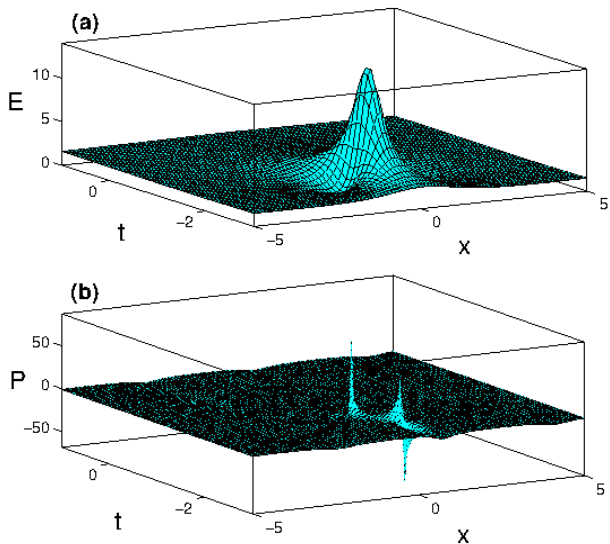}}
    	\caption{The energy density $\mathbf{E}$ (a) and the momentum $\mathbf{P}$ (b) fields in the $x-t$ plane for $A\sim a$ ---  Peregrine breather. The spatial period widens to present a colossal excitation localized in space 
    		and time.} 
    	\label{colossal}
\end{figure}

\begin{figure}[t]
	\resizebox{0.5\textwidth}{!}{\includegraphics[clip,trim=0cm 4cm 0cm 4cm,scale=1.0]{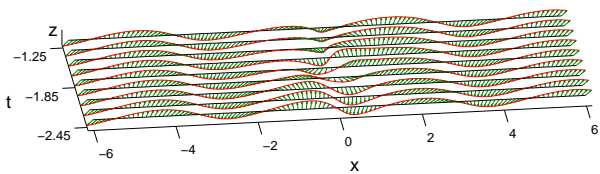}}
	\caption{The profile of the spin field in the narrow
		 $x-t$ region of rogue behavior. } 
	\label{profile}
\end{figure}

However, $|A-a|\sim 0$ is a case of special interest, as it is in
this regime that the spin field shows the rogue 
behavior. This can be understood from the 
expression for temporal period, \eq{period1}, when
$A<a$,
and similarly from \eq{period2} for the spatial 
period when $A>a$. Although we have already noticed glimpses 
of periodic rogue behavior in cases i and ii, the
phenomenon is more pronounced as $|A-a|\to 0$,  since both the periods
$T\,,L_2\,\to\infty$ (see Eqs. (\ref{period1} and \ref{period2},
respectively). 
Especially,
when $A>a$, a sudden localized colossal raise in
magnitude of the field is witnessed, as seen in the
plot of the energy density in Fig. \ref {colossal}.
A profile of the spin field itself for the narrow region
in space time when the rogue behavior is witnessed is shown in Fig. \ref{profile}.  An animation of the rogue
behavior of the spin field in illustrated in movie1. 

\section{Conclusion}
We have studied the breather spin configuration for 1-D
Heisenberg ferromagnet, whose dynamics is governed by the
integrable Landau-Lifshitz equation. The expression
for the breather field is explicitly obtained, and three 
distinct cases, based on relative parameter values studied.
Particularly, for a special choice of parameters, the breather shows a {\it rogue} behavior marked by a colossal
field excitation, yet local in space in time. As indicated in Section
2, the breather solutions correspond to a special periodic boundary
condition, which we believe is possible to obtain experimentally,
leading
to the  possibility of achieving a rogue spin field in the
laboratory.  The primary interest in solitons is due to their  characteristic properties --- localized, shape preserving, traveling and robust under collision. Of these, robustness under collisions has what has made them 
an attractive proposition application wise. Solitons owe
this property to the integrable nature of the dynamical
equation, and for this reason the same character must be 
carried over by rogue solitons as well under collision. Nevertheless, a direct confirmation on this aspect would 
open prospects of several applications based on the rogue behavior. In the near future, we intend to carry a detailed study on their collision behavior.

\end{document}